\documentclass{aa}  
\usepackage{graphicx}
\usepackage{txfonts}
\usepackage{natbib}
\begin{document}
   \title{The millimetre variability of M81*}

   \subtitle{Multi-epoch dual frequency mm-observations of the
   nucleus of M81}

   \author{R. Sch\"odel
          \inst{1}
          \and
          M. Krips
	  \inst{2}
          \and
          S. Markoff
	  \inst{3}
          \and
          R. Neri
	  \inst{4}
          \and
          A. Eckart
	  \inst{1}
          }

   \offprints{R. Sch\"odel}

   \institute{I. Physikalisches Institut, Universit\"at zu K\"oln,
              Z\"ulpicher Str. 77, D-50937 K\"oln, Germany\\
              \email{rainer@ph1.uni-koeln.de,eckart@ph1.uni-koeln.de}
         \and
         Harvard-Smithsonian Center for Astrophysics, SMA project, 645
              North A'ohoku Place, Hilo, HI 96720\\
         \email{mkrips@cfa.harvard.edu}
         \and
         Sterrenkundig Instituut ``Anton Pannekoek '', Universiteit
          van Amsterdam, Kruislaan 403, 1098 SJ Amsterdam, The Netherlands\\
          \email{sera@science.uva.nl}
          \and
          Institut de Radio Astronomie Millim\'etrique, 300 rue de la
              Piscine, Domaine Universitaire, 38406 Saint Martin
              d'H\`eres, France\\
          \email{neri@iram.fr}
            }

   \date{Received September ; accepted }

% \abstract{}{}{}{}{} 
% 5 {} token are mandatory
 
  \abstract
  % context heading (optional)
  % {} leave it empty if necessary  
   {}
  % aims heading (mandatory) 
 {There are still many open questions as to the physical mechanisms at
  work in Low Luminosity AGN that
  accrete in the extreme sub-Eddington regime. Simultaneous
  multi-wavelength studies have been very successful in constraining
  the properties of Sgr~A*, the extremely sub-Eddington black hole at
  the centre of our Milky Way. M81*, the nucleus of the nearby spiral galaxy
  M81, is an ideal source to extend the insights obtained on Sgr~A*
  toward higher luminosity AGN. Here we present observations at 3 and
  1\,mm that were obtained within the framework of a coordinated,
  multi-wavelength campaign on M81*.} 
 % methods heading (mandatory) 
 {The continuum emission from M81* was observed during three epochs with the IRAM Plateau de Bure
  Interferometer simultaneously at wavelengths of 3 and 1\,mm.}  
% results heading   (mandatory)   
{ We present the first flux measurements of M81* at wavelengths
 around 1\,mm.  We find that M81* is a continuously variable source
 with the higher variability observed at the shorter wavelength. Also,
 the variability at 3 and 1\,mm appears to be correlated. Like Sgr~A*,
 M81* appears to display the strongest flux density and variability in
 the mm-to-submm regime. There remains still some ambiguity concerning
 the exact location of the turnover frequency from optically thick to
 optically thin emission.  The observed variability time scales point
 to an upper size limit of the emitting region of the order
 25~Schwarzschild radii.}
% conclusions  heading 
{ The data show that M81* is indeed a system with very similar
  physical properties to Sgr~A* and an ideal bridge toward high
  luminosity AGN. The data obtained clearly demonstrate the usefulness
  and, above all, the necessity of simultaneous multi-wavelength
  observations of LLAGN.}

   \keywords{LLAGN --
                mm-observations --
                galactic nucleus
               }

   \maketitle
%
%________________________________________________________________

\section{Introduction}

 At high accretion rates onto black holes, the infall of gas and
dissipation of energy via mostly thermal processes in a thin disk are
fairly well understood in general terms.  In contrast, there is still
great debate about which physical mechanisms are dominant in objects
which are accreting at well below the Eddington rate.  The most
extremely sub-Eddington source currently accessible to observations
that allows still reasonable statistics for modelling the
accretion/emission mechanisms is the central supermassive black hole
of our Galaxy, Sagittarius~A* (Sgr~A*).  Its low luminosity has
perplexed theorists for decades, and stimulated the re-emergence of
radiatively inefficient accretion flow models \citep[RIAFS,
see][]{Melia2001ARA&A,Quataert2003ANS}, including the advection
dominated accretion flow model \citep[ADAF, e.g.\
][]{NarayanYi1994ApJ} and its derivatives.  These models are
characterised by low efficiency in converting thermal energy into
electromagnetic radiation, but differ in terms of the inner boundary
conditions such as the presence of strong winds \citep[e.g.\
][]{BlandBegel1999MNRAS} and/or convection
\citep{QuataertGruzinov1999ApJ}, each of which lead to differing
predictions on the detailed spectral energy distribution (SED) and its
variability.  Another open question is the role of jets in dissipating
accretion energy \citep[e.g.\
][]{BlandKonigl1979ApJ,FalckeBiermann1995A&A} and the extent of their
contribution to the SED, particularly at higher frequencies
\citep[e.g.\
][]{FalckeMarkoff2000A&A,Markoff2001A&A,Markoff2001A&Ab,Yuan2002A&A}.
Since jets are observed in almost all accreting astrophysical systems,
the importance of jets in weakly accreting systems is a matter of
particular interest.
 
With a luminosity $\sim$10$^{-9\ldots-10}\times$L$_{\rm Edd}$, Sgr~A*
is -- in terms of Eddington luminosity -- the weakest accreting black
hole with observational statistics good enough to fit models to its
spectrum.  It is the primary testbed for theoretical models of extreme
sub-Eddington accretion, which rely largely on the available
radio/submm/NIR/X-ray observations of Sgr~A* \citep[for a review
see][]{Melia2001ARA&A}.  The recent discovery of quiescent and flaring
emission from Sgr~A* at both X-ray and NIR wavelengths
\citep{Baganoff2001Natur,Genzel2003Natur,Ghez2004ApJ} has
revolutionised our understanding of this particular Low Luminosity
Active Galactic Nucleus (LLAGN).  The stringent constraints from these
data have ruled out several models, leaving only RIAF models, jet
models, and combinations thereof as contenders. Recently,
\citet{Eckart2004A&A,Eckart2006A&A} were successful in obtaining the
first simultaneous measurements of the emission from Sgr~A* at
X-ray/NIR wavelengths (with quasi-simultaneous sub-mm
observations). These and ongoing coordinated multi-wavelengths
campaigns deliver the decisive observations for
constraining/eliminating weak accretion models further.

The nearby spiral galaxy M81 (NGC~3031) is an Sb spiral galaxy similar
to the Milky Way. It is located at a distance of $3.63\pm0.34$\,Mpc
\citep{Freedman1994ApJ}. \citet{Devereux2003AJ} used spectroscopic
measurements of the H$\alpha+[$N$II]$ emission, probably emitted from
a rotating gas disk inclined at an angle of $14^{\circ}\pm2^{\circ}$,
to infer a mass of $7.0^{+2}_{-1}\times10^{7}$\,M$_{\odot}$ for the
central black hole in M81.

The nucleus of M81, termed M81*, shows typical signs of AGN
activity. It has a power-law, variable X-ray continuum
\citep{Ishisaki1996PASJ,Page2004A&A}. The X-ray flux from M81 is
highly variable, at scales from days to years \citep{LaParola2004ApJ}.
The nucleus of M81 displays double peaked, broad H$\alpha$ emission
lines \citep{Bower1996AJ}. However, the overall luminosity and AGN
characteristics of M81* are rather weak and the galaxy is classified
as a LINER \citep[low-ionisation nuclear emission-line region,
e.g.\ ][]{Ho1996ApJ}. With a luminosity of the order of $10^
{37}$\,erg\,s$^{-1}$ in the radio and $10^ {40}$\,erg\,s$^{-1}$ in the
optical/X-ray domains \citep[see, e.g., compilations
by][]{Ho1996ApJ,Ho1999ApJ}, its luminosity is $<10^{-5}$ times the
Eddington luminosity in any wavelength regime.  M81* is therefore
counted among the low-luminosity AGN (LLAGN).  It shows the typical
spectral energy distribution (SED) of this class of sources, that is
characterised by the absence of the so-called big blue bump, the
ultraviolet excess found commonly in the higher power AGN \citep[e.g.\
][]{Ho1999ApJ}.

At cm-wavelengths, M81* shows large-amplitude variations (factors up
to two at 2\,cm) with timescales of a few months and weaker changes of
the flux density on timescales $\leq1$~day
\citep{Ho1999AJ}. Multi-epoch VLBI observations of M81* at
$\sim$$0.01$\,pc resolution at 20 epochs over $4.5$\,yr reveal a
stationary core with a variable (on timescales of $\sim$1\,yr)
one-sided jet of length 1\,mas (3600\,AU) towards the northeast
\citep{Bietenholz2000ApJ}. As for the polarisation properties of M81*,
circular polarisation was detected at $4.8$, $8.4$
\citep{Brunthaler2001ApJ}, and 15\,GHz \citep{Brunthaler2006A&A},
while linear polarisation appears to be absent at these
frequencies. This is an intriguing similarity to Sgr~A*, where
circular polarisation also dominates over linear polarisation at these
wavelengths
\citep[][]{Bower1999ApJa,Bower1999ApJb,Bower1999ApJc,Aitken2000ApJ,Bower2003ApJ,Bower2005ApJ}.

\citet{Sakamoto2001AJ} present observations of the central kiloparsec
of M81 at a wavelength of 3\,mm in the CO $J=1-0$ line and continuum
at 100\,pc resolution. They detect molecular gas in a pseudoring or
spiral arm at about 500\,pc, but no giant molecular cloud within about
300\,pc of the nucleus. They find significant intraday variation of
the continuum emission from M81*, suggesting an emitting region of
$\sim$100\,AU.

\citet{ReuterLesch1996A&A} obtained a spectrum of M81* from the radio
to the mm-regime. They find an inverted spectrum up to 100\,GHz. Its
flux density can be described well by the law $S_{\nu} \propto
\nu^{1/3}\exp{-(\nu/\nu_{c})}$ with a turnover frequency of $\nu_{c} =
200$\,GHz. They point out the remarkable similarity between the
spectrum of M81* and Sgr~A* and conclude that the same physical
mechanisms might operate in both galactic nuclei. Therefore
\citet{ReuterLesch1996A&A} suggest that M81* may be a by
$\sim$10$^{4}$ upscaled version of Sgr~A*.

Thus M81* is a unique source for comparison with Sgr~A* and more
powerful AGN, and constitutes the next logical step after the
successful multi-wavelength observations of Sgr~A*. With this aim, a
coordinated, multi-wavelength campaign took place in the first half of
2005, involving instruments from the X-ray to the radio domain: the
Chandra X-ray observatory, the Lick telescope (NIR), the SMA, the
Plateau de Bure Interferometer (PdBI), the VLA, and the GMRT. A
compilation of all observations and the interpretation of the
multi-wavelength data is presented by Markoff et al.\ (in
preparation).  A detailed description of all instruments and the
related data reduction would overload the multi-wavelength
paper. Therefore, some papers are dedicated to the observations with
specific instruments.  The X-ray emission lines measured in the
Chandra observations are discussed by Young et al.\ (in
preparation). In this paper we focus on the three epochs of
mm-observations of M81* that were obtained with the PdBI during this
campaign.

%__________________________________________________________________

\section{Observations and data reduction}

%_____________________________________________________________
%
\begin{table*}
\caption{Observations of M81* with the PdBI during 2005.  $\nu_{\rm
    3\,mm}$ is the exact frequency used around 3\,mm, $\nu_{\rm
    1\,mm}$ the one at a wavelength of 1\,mm. N$_{\rm Ant}$ is the
    number of antennae, ``Config'' refers to the antenna configuration
    used. }
\label{table:obs}      
\centering          
\begin{tabular}{cccccc}     % 7 columns 
\hline\hline       
Start [UT] & End [UT] & $\nu_{\rm 3\,mm}$ [GHz] &  $\nu_{\rm 1\,mm}$ [GHz] & N$_{\rm Ant}$ & Config\\
\hline                    
24-FEB-2005~01:11 & 24-FEB-2005~19:45 & 115.3 & 230.5 & 6 & Bp \\
14-JUL-2005~06:50 & 15-JUL-2005~13:51 & 80.5 & 241.4 & 5 & D \\
19-JUL-2005~23:17 & 20-JUL-2005~16:07 & 86.2 & 218.2 & 5 & D \\
\hline                  
\end{tabular}
\end{table*}
%
%_____________________________________________________________
%_____________________________________________________________
%
\begin{table*}
\caption{Data quality: The table lists the ranges of the rms values of
  the phase and amplitude between all baselines for each observing epoch
  and wavelength. Also, the estimated systematic error of the absolute
  flux calibration is indicated.}
\label{Tab:quality}      
\centering          
\begin{tabular}{l|c|c|c|c||c|c}
\hline\hline       
     & 3\,mm & 3\,mm & 3\,mm & 1\,mm & 1\,mm & 1\,mm \\
Date & phase rms [deg] & amp rms [\%] & absolute & phase rms [deg] &
     amp rms [\%] & absolute \\
\hline                    
24-FEB-2005    & 10-20 & 5-8     & $\leq$10\% & 20-40 & 11-18 &
     $\leq$15\% \\
14/15-JUL-2005 & 15-35 & 8-11    & $\leq$15\% & 40-60 & 25-30 &
     $\leq$30\% \\
19/20-JUL-2005    & 12-28 & $\sim$3 & $\leq$15\% & 30-70 & 11-18 &
     $\leq$20\% \\
\hline                  
\end{tabular}
\end{table*}
%
%_____________________________________________________________

M81* was observed with the PdBI on 24 February, 14-15 July, and 19-20
July 2005. Six antennae were used for the February observations in the
B configuration of the PdBI, which provides typical beam sizes around
$1.5''$ at $\sim$100\,GHz and around $0.8''$ at $\sim$200\,GHz. The
July observations were done with five antennae in the more compact D
configuration, with typical beam sizes around $5''$ at $\sim$100\,GHz
and $2.5''$ at $\sim$200\,GHz. All observations are listed in
Table~\ref{table:obs}.

The observations aimed to detect continuum emission from M81*.
However, due to good conditions the receivers were tuned to the
$^{12}$CO $J=1-0$ and $J=2-1$ transitions at $115.3$ and $230.5$~GHz
in the February observations. Thus, it was possible to search for
compact CO-emission within $\sim$20$''$ of the nucleus, while the
continuum could be extracted from the line-free channels (The results
of the CO-line imaging will be discussed in a forthcoming paper).
For greater phase stability in the July observations, an epoch during
which the atmosphere at the site contains more water vapour, the
receivers were tuned to frequencies of $80.5$ and $86.2$\,GHz at 3\,mm
and $241.4$ and $218.2$\,GHz at 1\,mm.

The sources 3C273 (February), 3C454.3 (July 14-15) and 1741-038 (July
19-20) were used for bandpass calibration. Phase calibration was
performed with the sources 1044+719 and 0836+710. Primary flux
calibrators to determine the efficiencies of the antennae were the
sources 1044+719 for February 23-24 (1.6\,Jy at 3\,mm/1.1\,Jy at
1\,mm), 1044+719 (1.8\,Jy at 3\,mm) and 2200+420 (8.7\,Jy at 1\,mm)
for July 14/15, and MWC349 (1.0\,Jy at 3\,mm) and 3C$454.3$ (33.0\,Jy
at 1\,mm) for July 20.  The phase calibrators 1044+719 and 0836+710
were used to fit the time-dependent fluctuations of the amplitude for
all baselines.  Various tests were performed for estimating the
uncertainty of the flux calibration, e.g.\ using different primary
flux calibrators and comparing the resulting fluxes of all observed
sources. As an additional test, from a comparison of the calibrated
fluxes of the sources 0836+710, 1044+719, and MWC349 between the three
epochs we estimated the uncertainty of the absolute flux
calibration. To provide a quantitative measurement of the
data quality, Table~\ref{Tab:quality} lists the ranges of the rms
values of the phase and amplitude that were obtained during
calibration of these values for the different baselines for each
observing epoch. The table also provides values for the estimated
uncertainty of the absolute flux calibration at 3\,mm and 1\,mm.  As
can be seen, the quality of the February data is highest. The data
from 20 July are clearly better than the data from 14/15 July.

%------------------------------------
\begin{figure*}
  \includegraphics[height = .8\textheight]{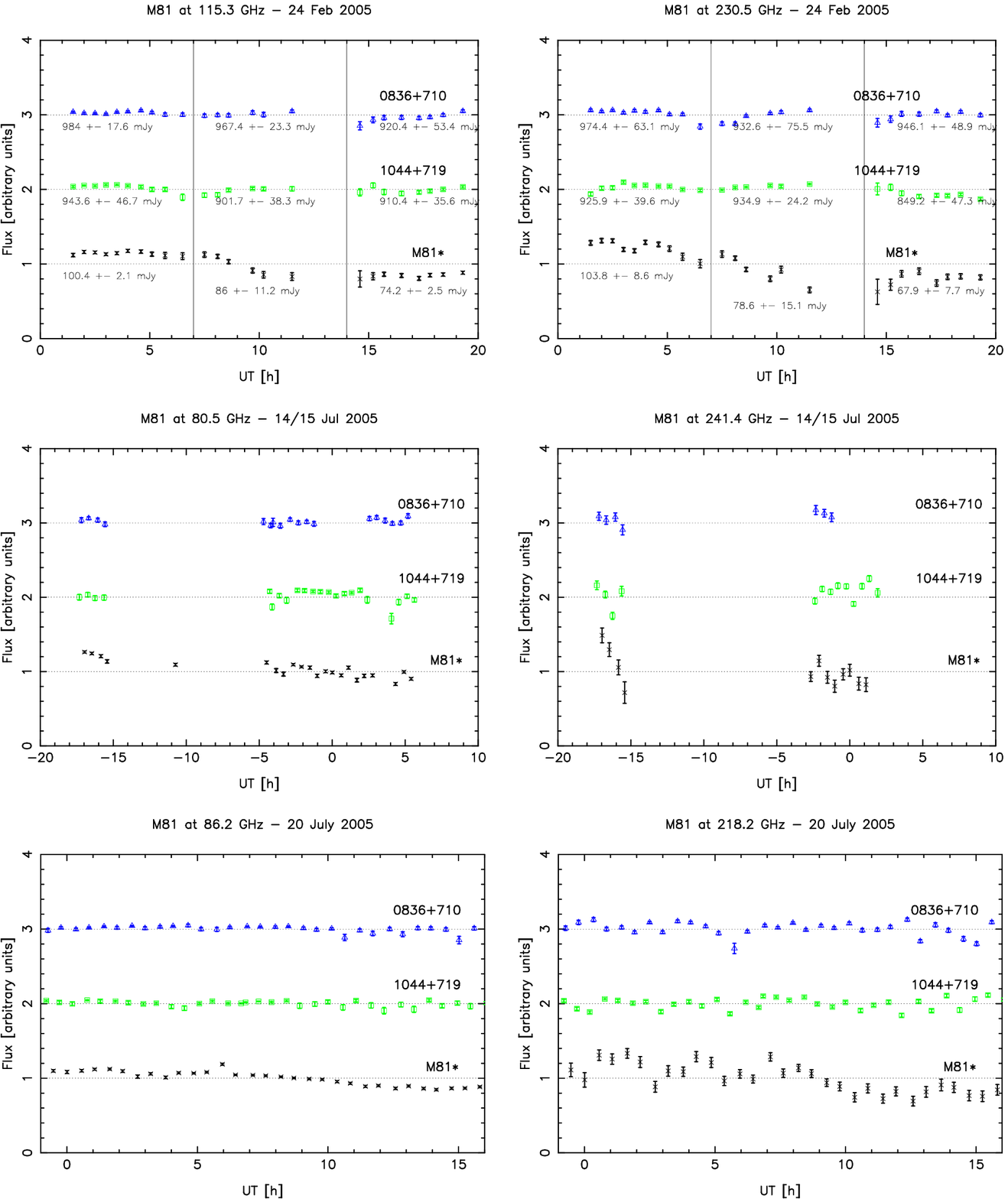}
  \caption{Light curves of M81* and of the calibrators 1044+719 and
  0836+710 from the PdBI observations at 3 and 1\,mm. The x-axes show
  UT in hours. The fluxes were all scaled to an average value of 1,
  with the curves of the calibrators shifted for better comparison.
  The vertical lines in the top panels indicate sections for which the
  average flux and the standard deviation of the individual
  measurements have been calculated. The corresponding values are
  indicated. \label{Fig:var}}
\end{figure*}
%-------------------------------------

%------------------------------------
\begin{figure*}
  \includegraphics[width=\textwidth]{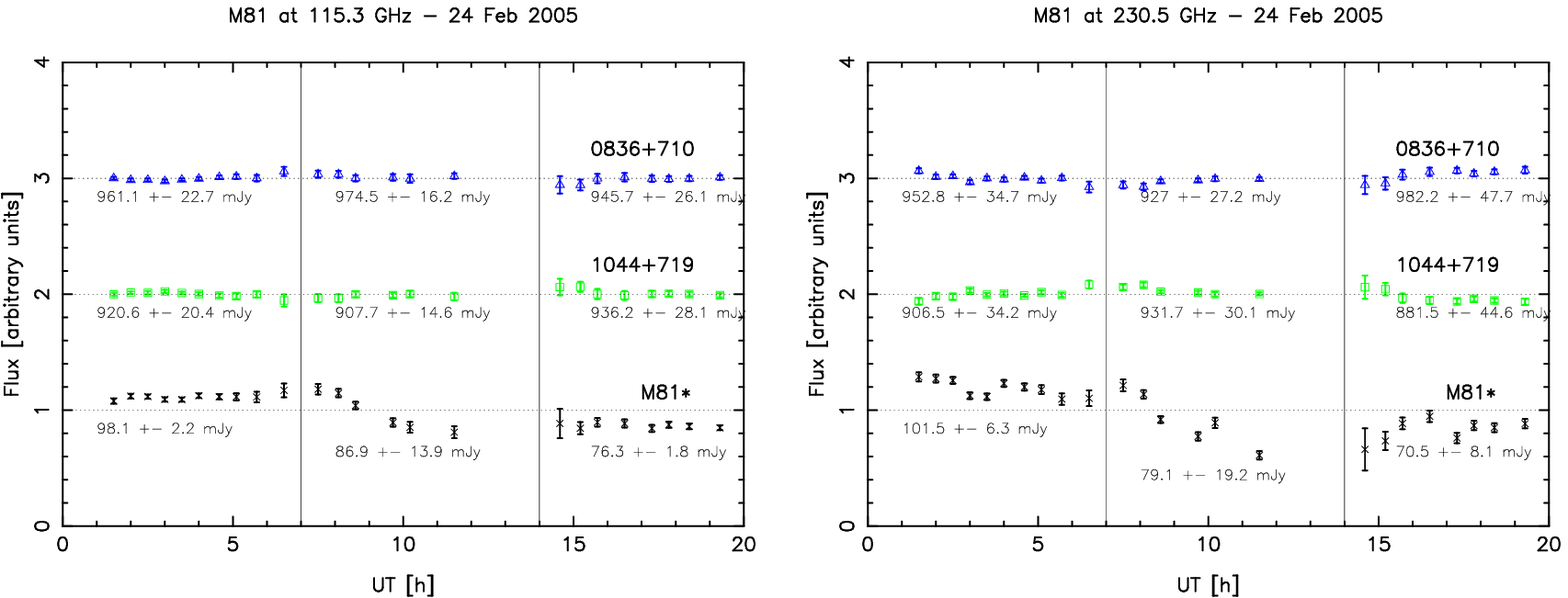}
  \caption{ Light curves of M81* and of the calibrators 1044+719 and
  0836+710 from the PdBI observations at 3 and 1\,mm on February
  24. The light curves were corrected for possible systematic
  variations by forcing the calibrator sources to have constant flux
  (on average).
 \label{Fig:calib}}
\end{figure*}
%-------------------------------------

Individual scans of 20~min duration were extracted from the calibrated
data. Subsequently, a point source was fitted to the resulting UV
tables in order to determine the flux of M81* and of the two
calibrators 0836+710 and 1044+719. Fitting Gaussian functions to the
UV tables of M81 gave very similar results, with deviations $\leq5\%$
at 3\,mm and $\leq15\%$ at 1\,mm. Due to the larger number of
parameters the uncertainties are larger when fitting Gaussians. The
general shape of the radio light curves is not affected by the model
that is chosen to fit the data. Since M81* is known to be a point-like
radio continuum source that shows structure only on scales of
1\,milli-arcsecond \citep{Bietenholz2000ApJ}, we decided that fitting
a point source appears to be the appropriate way to measure the flux density
of M81*.

%_____________________________________________________________
%
\begin{table*}
\caption{Average flux densities of M81*, $S_{\rm 3\,mm}$ and $S_{\rm
   1\,mm}$, measured at $\nu_{\rm 3\,mm}$ and $\nu_{\rm 1\,mm}$ during
   the three epochs of the campaign.  The mean and standard deviation
   (not the error of the mean) were calculated from the unweighted
   individual measurements shown in Fig.~\ref{Fig:fluxes}. The last
   column lists the average spectral indices and their standard
   deviation (not the error of the mean) calculated from all
   individual measurements as shown in the right panels of
   Fig.~\ref{Fig:index}.}
\label{table:meanfluxes}      
\centering          
\begin{tabular}{ccccccc}
\hline\hline       
Start [UT] & End [UT] & $\nu_{\rm 3\,mm}$ [GHz] & $S_{\rm 3\,mm}$ & $\nu_{\rm 1\,mm}$ [GHz] &  $S_{\rm 1\,mm}$ & $\alpha$ \\
\hline                    
24-FEB-2005~01:11 & 24-FEB-2005~19:45 & 115.3 & $88.0\pm11.7$ & 230.5 & $85.6\pm17.8$ & $-0.06\pm0.16$ \\
14-JUL-2005~06:50 & 15-JUL-2005~13:51 & 80.5 & $241.2\pm33.8$ & 241.4 & $181.2\pm39.1$ & $-0.38\pm0.17$ \\
19-JUL-2005~23:17 & 20-JUL-2005~16:07 & 86.2 & $118.7\pm11.4$ & 218.2 & $74.8\pm13.3$ & $-0.51\pm0.15$ \\
\hline                  
\end{tabular}
\end{table*}
%
%_____________________________________________________________

\section{Variability of M81* at mm wavelengths}

Figure~\ref{Fig:var} shows the observed millimetre light curves of
M81* along with the corresponding data for the two phase calibrators.
The scaling has been chosen such that the average flux of each source
equals a value of 1 (arbitrary units), with the fluxes of the
calibrators shifted for better comparison.  Only the error bars of the
relative uncertainty are shown. As for the absolute uncertainty of
calibration, it would merely shift the curves along the vertical axes
and have no influence on the shape of the curves under stable
observing conditions. The flux of the calibrator sources should then
fluctuate randomly and with small standard deviation around a constant
value.  This is the case for the 3\,mm-data on 24 February and 20
July, while the conditions were less favourable on 14/15 July. In case
of the 1\,mm-data, the calibrator measurements fluctuate around a
stable average on 20 July, they show trends in parts of the data from
24 February (that are, however, significantly smaller and
significantly less pronounced than the trend seen in the flux from
M81*), and show significant deviations on 14/15 July.

As for the overall quality of the data, it can be best asserted when
examining the calibrator sources (see also Tab.~\ref{Tab:quality} for
more quantitative information on the data quality). Clearly, the
quality of the data is higher at 3\,mm than at 1\,mm due to the
reduced phase stability at the shorter wavelength. The best data were
obtained on February 24, when the lower winter temperatures and vapour
content of the atmosphere facilitated measurements at 1\,mm. The July
20 observations are excellent at 3\,mm and still good at
1\,mm. Observing conditions were highly variable on 14/15 July, but it
was still possible to obtain good data at 3\,mm and acceptable ones at
1\,mm. However, a large part of the data had to be discarded during
calibration so that in total only about 6~hours of usable data were
left.

\begin{figure*}
  \includegraphics[width =\textwidth]{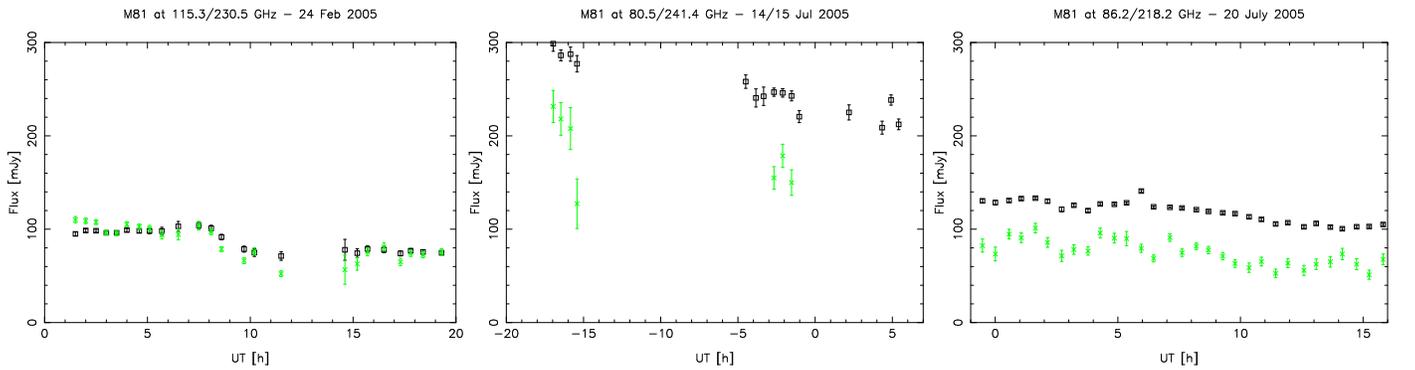}
  \caption{Flux density of M81* on 24~February, 14/15~July, and
  20~July 2005. The black boxes mark the flux density at 3\,mm and the
  green (gray) crosses mark the flux density at 1\,mm. The exact
  frequencies are indicated in the titles of the individual
  panels. The error bars indicate relative, not absolute,
  uncertainties.
\label{Fig:fluxes}}
\end{figure*}

\begin{figure*}
  \includegraphics[width =\textwidth]{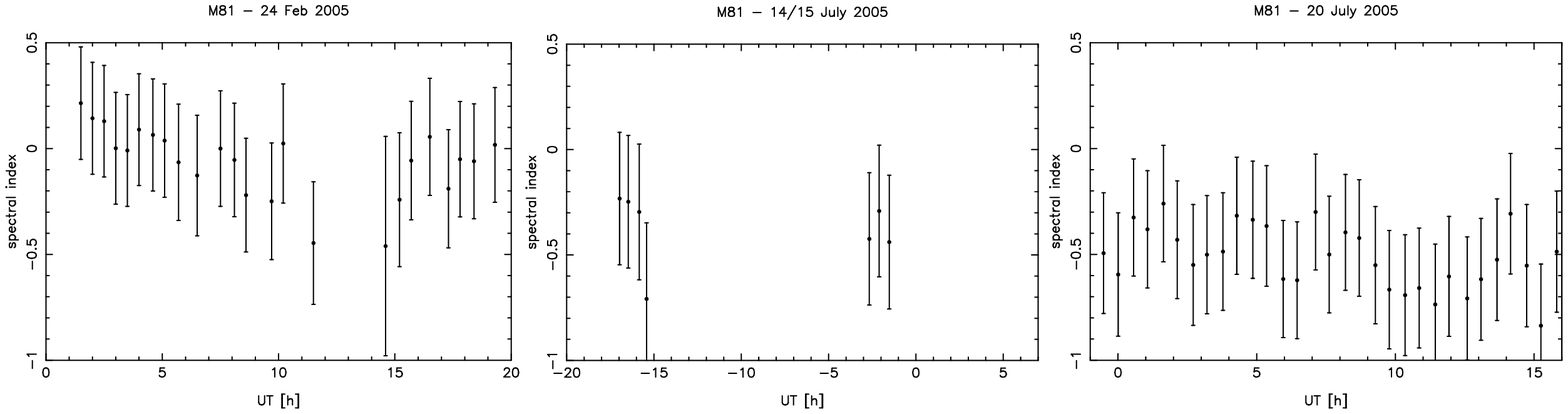}
  \caption{ Spectral index of M81* on 24~February, 14/15~July, and
  20~July 2005. The spectral index is defined by
  $S_{\nu}\propto\nu^{\alpha}$, where $S_{\nu}$ is the flux density at
  a given frequency $\nu$. The absolute uncertainties (see
  Tab.~\ref{Tab:quality}) of the flux densities at 3 and 1\,mm have
  been taken into account for this plot. Since we use upper limits on
  the absolute uncertainties, the plotted error bars are conservative.
\label{Fig:index}}
\end{figure*}

Some variability of the calibrators can be seen in Fig.~\ref{Fig:var}
that may be correlated with the variability of the target. We tried to
take this possibility into account by introducing an additional
calibration step that forces the flux of the calibrators to be
constant (on average). Thus, any possible systematic uncertainties,
leading to spurious variability, can be
removed. Figures~\ref{Fig:calib} to \ref{Fig:index} and all values in
Tab.~\ref{table:meanfluxes} refer to measurements after this final
calibration step was applied (uncertainties introduced by the
calibration process were taken into account).  Generally, the effect
of this additional calibration step is minor. It is illustrated for
the Feb.~24 data in Fig.~\ref{Fig:calib}. Since (quasi)-simultaneous
measurements of sufficient quality for both calibrators were not
available for all measurements of M81*, there are less data points in
some of the plots in Fig.~\ref{Fig:fluxes} and \ref{Fig:index} than in
Fig.~\ref{Fig:var}.

The calibrated flux density measurements of M81* at 3\,mm and 1\,mm
are shown in Fig.~\ref{Fig:fluxes} (only error bars of the relative
uncertainty are shown) and the corresponding spectral indices in
Fig.~\ref{Fig:index} (here, the absolute uncertainties of the flux
calibration have been taken into account additionally).  The spectral
index is defined by $S_{\nu}\propto\nu^{\alpha}$, where $S_{\nu}$ is
the flux density at a given frequency $\nu$. Time averages of the flux
densities and spectral indices are listed in
Table~\ref{table:meanfluxes}.

We find the following characteristics of the variability of M81* at
wavelengths of 3\,mm and 1\,mm:
\begin{itemize}
\item Variability at the two wavelengths appears to be correlated.
\item The amplitude of variability is by a factor $\sim$1.5 larger
  at 1\,mm.
\item On 14 July, the flux density at 3\,mm is a factor of $\sim$3, at
  1\,mm a factor of $\sim$2 higher than at the other two epochs.
\item The spectral index varies between $\sim$0 and $\sim$-0.5
  (Fig.~\ref{Fig:index} and Tab,~\ref{table:meanfluxes}).
\item M81* shows intraday variability, with the fastest variability
  observed between 08\, and 12\,h UTC on 24 February.
\end{itemize}

The light curve of February 24 is shown in the upper panels of
Fig.~\ref{Fig:var}. The average flux density and its standard
deviation is indicated for different sections of the light curve. A
$\gtrsim5\sigma$ decrease can be seen at 3\,mm between 7 and 12~UT. It
is accompanied by a similar decrease at 1\,mm. The February 24 light
curves are shown in Fig.~2 after forcing the two calibrator sources to
have a constant flux (on average). This additional calibration step
does not alter the intrinsic shape of the light curves in a
significant way. Therefore the observed intra-epoch variability must
have been intrinsic to the source.

An important question is whether the detected variability could be
related to polarisation.  We can discard this possibility, however,
because a) The directions of linear polarisation of the 3 and 1\,mm
receivers at the antennae of the PdBI are orthogonal. If a sinusoidal
pattern due to polarisation were observed at 3\,mm, one would expect
to observe a similar pattern at 1\,mm, shifted by several hours. We do
not detect the expected sinusoidal pattern in the light curves nor are
the light patterns shifted by several hours in time relative to each
other.  b) Recent observations of M81* with the BIMA array by
\citet{Brunthaler2006A&A} indicate absence of or very low upper limits
of a few percent on linear and circular polarisation of M81* at 86 and
230\,GHz.
%-------------------------------------

%-------------------------------------
%-------------------------------------

%-------------------------------------

\section{Discussion}

Theory and observations indicate that sub-Eddington black holes are
jet-dominated
\citep{FalckeMarkoff2000A&A,Falcke2004A&A,Yuan2002A&A,Fender2003MNRAS}.
A compact, variable jet has in fact been detected in VLBI observations
of M81* (Bietenholz et al., 2000; see also Markoff et al., in
preparation). \citet{Falcke1996ApJ} demonstrate that a compact jet can
explain the observed spectrum of M81* very well.  In the jet model,
the turnover frequency from a flat (or slightly inverted) radio
spectrum to an optically thin power-law occurs at a frequency
$\nu_{t}$. In a simplified model this frequency depends on the jet
power, $Q_{j}$, and mass of the black hole, $M_{\rm BH}$, as
$\nu_{t}\propto Q_{j}^{2/3}M_{\rm BH}^{-1}$ \citep{Falcke2004A&A}.  In
LLAGN the turnover occurs generally at (sub)mm wavelengths, while it
is located in the near-infrared/optical regime for XRBs in the
low/hard state
\citep[see][]{CorbelFender2002ApJ,Markoff2003A&A,Falcke2004A&A}.

The simultaneous observations of M81* at 3 and 1\,mm presented in this
work are consistent with a turnover of the synchrotron emission from a
jet in M81* into the optically thin part between 3\,mm and 1\,mm (with
the exception of the first $\sim$4~hours of the 24~February light
curve that would indicate a higher turnover frequency).  This is in
good agreement with the data presented by \citet{ReuterLesch1996A&A}
that were, however, not acquired simultaneously.

The evidence for the location of the turnover frequency between 3 and
1\,mm is ambiguous, however.  The multi-frequency radio data from the
coordinated campaign indicate that bumps are present in the radio
spectrum (Markoff et al., in prep.). These bumps are located at
different frequencies at different epochs. Therefore, in spite of the
presented mm-data, it may well be possible that there is a submm-bump
present in M81*. Observations with the SMA at 345\,Ghz that were
obtained during the coordinated campaign on M81* indicate that the
flux density of M81* increases toward the submm regime, in agreement
with theoretical predictions on the existence of a submm-bump (Markoff
et al., in preparation).  Unfortunately, there was only one epoch (24
February 2005), where measurements with the PdBI and the SMA were
simultaneous.

For fine-tuning the models of the emission of M81* it is important
to know the exact turnover frequency and whether and how it varies
with activity of the source.  Therefore there is an urgent need for
more observational data for comparison with theory in order to
understand sub-Eddington accretion and emission. It is essential that
the observations across the wavelength regimes are simultaneous.

Due to the gap of several months there is probably no correlation
between the February and July observations because the light curves at
2\,cm presented by \citet{Ho1999AJ} indicate that radio outbursts of
M81* have generally timescales $<5$\,months. Also,
\citet{Sakamoto2001AJ} have found that M81* shows flux variations {\bf
of factors $\leq2$} at a wavelength of 3\,mm on time scales
$<10$\,days.  However, the decaying light curve from July 19 may be
related to the same event as the light curve on July 14. The spectral
index during the two epochs is negative and of similar absolute value
(see Tab.~\ref{table:meanfluxes} and Fig.~\ref{Fig:index}). This
behaviour is consistent with a decaying light curve from a low-peaking
flare in the model by \citet{Valtaoja1992A&A} that describes
variability phenomena in AGN by a growth/decay of shocks in a jet
\citep[see also][]{MarscherGear1985ApJ}.

An estimate of the size scales of the relevant processes can be
obtained from the time scales of the observed variability. The
drop-off between 07:00\,UT and 12:00\,UT in the highest quality data
set from 24 February 2006 shows that a $\gtrsim5\sigma$ variability of
the flux at 3\,mm occurs on a time scale of 5~hours \citep[see also
the intraday variability at 3\,mm found by ][]{Sakamoto2001AJ}. A
similarly rapid change of the flux can be seen at 1\,mm.  Since no
signal can travel faster than at the speed of light this corresponds
to an upper limit on the size of the source of merely $\sim$25~
Schwarzschild radii, when a mass of $7.0\times10^{7}$\,M$_{\odot}$ is
assumed for the black hole in M81*.

\section{Summary}

We present three epochs of simultaneous 1 and 3\,mm continuum
observations of the LLAGN M81* that were obtained in the framework of
a coordinated, multi-wavelength campaign (Markoff et al., in
preparation).

The observations of M81* with the PdBI at mm-wavelengths confirm that
M81* is a continuously varying radio continuum source as has been
found previously, mostly at larger wavelengths \citep[see
][]{Ho1999AJ,Bietenholz2000ApJ,Sakamoto2001AJ}. The measurements
present the first unambiguous detection of M81* at 1\,mm and moreover
show that the source is continuously variable at this wavelength as
well.  The amplitude of the variability is observed to be generally
larger at 1\,mm than at 3\,mm by a factor of roughly 1.5. This agrees
well with the trend found by \citet{Ho1999AJ} that the amplitude of
the variability increases with frequency in LLAGN.  A similar
behaviour is found in the far weaker source Sgr~A* \citep[e.g.\
][]{Herrnstein2004AJ,Miyazaki2004ApJ}.

The shortest variability time scales of our observations give upper
limits on the size of the emitting region of $\sim$25\,Schwarzschild
radii, assuming a black hole mass of $7.0\times10^{7}$\,M$_{\odot}$.

The decaying light curves observed on 14/15 and on 20 July 2005 may
have been related to the same radio outburst. They are consistent with
a generalized shock-in jet model \citep{Valtaoja1992A&A}.

The simultaneous measurements of M81* at 3 and 1\,mm are consistent
with a turnover of the flux in the mm-to-submm regime as predicted by
models for a jet dominated source.  Some ambiguity remains concerning
the exact peak frequency and the related interpretation of variability
events.

The observations confirm previous findings that there are many
similarities between M81* and Sgr~A*, the source related to the
supermassive black hole at the center of the Milky Way. This
underlines the similarity between LLAGN despite of several orders of
magnitudes of difference between their luminosities. M81* can serve as
a bridge from the extremely sub-Eddington Sgr~A* toward higher
luminosity LLAGN. The radio emission from M81* is apparently dominated
by a compact jet. The same may be the case for Sgr~A*, although such a
jet has not yet been discovered due to the strong interstellar
scattering toward this source. This work poses a clear case for
follow-up observations of the highly sub-Eddington nucleus M81* and
for the necessity of simultaneous multi-wavelength observations in
general in order to constrain the physical mechanisms in black holes
that accrete at rates several orders of magnitude below the Eddington
limit.

\begin{acknowledgements}
Part of this work was supported by the \emph{Deutsche
Forschungsgemeinschaft, DFG} project number 494. We thank Lena Lindt
and Jan Martin Winters for their help with the data reduction. Special
thanks to M.~Nowak fo his efforts in setting up the coordinated
campaign on M81.

\end{acknowledgements}

\bibliography{rainer}

\begin{thebibliography}{45}
\expandafter\ifx\csname natexlab\endcsname\relax\def\natexlab#1{#1}\fi

\bibitem[{{Aitken} {et~al.}(2000){Aitken}, {Greaves}, {Chrysostomou},
  {Jenness}, {Holland}, {Hough}, {Pierce-Price}, \& {Richer}}]{Aitken2000ApJ}
{Aitken}, D.~K., {Greaves}, J., {Chrysostomou}, A., {et~al.} 2000, \apjl, 534,
  L173

\bibitem[{{Baganoff} {et~al.}(2001){Baganoff}, {Bautz}, {Brandt}, {Chartas},
  {Feigelson}, {Garmire}, {Maeda}, {Morris}, {Ricker}, {Townsley}, \&
  {Walter}}]{Baganoff2001Natur}
{Baganoff}, F.~K., {Bautz}, M.~W., {Brandt}, W.~N., {et~al.} 2001, \nat, 413,
  45

\bibitem[{{Bietenholz} {et~al.}(2000){Bietenholz}, {Bartel}, \&
  {Rupen}}]{Bietenholz2000ApJ}
{Bietenholz}, M.~F., {Bartel}, N., \& {Rupen}, M.~P. 2000, \apj, 532, 895

\bibitem[{{Blandford} \& {Begelman}(1999)}]{BlandBegel1999MNRAS}
{Blandford}, R.~D. \& {Begelman}, M.~C. 1999, \mnras, 303, L1

\bibitem[{{Blandford} \& {Konigl}(1979)}]{BlandKonigl1979ApJ}
{Blandford}, R.~D. \& {Konigl}, A. 1979, \apj, 232, 34

\bibitem[{{Bower} {et~al.}(1996){Bower}, {Wilson}, {Heckman}, \&
  {Richstone}}]{Bower1996AJ}
{Bower}, G.~A., {Wilson}, A.~S., {Heckman}, T.~M., \& {Richstone}, D.~O. 1996,
  \aj, 111, 1901

\bibitem[{{Bower} {et~al.}(1999{\natexlab{a}}){Bower}, {Backer}, {Zhao},
  {Goss}, \& {Falcke}}]{Bower1999ApJa}
{Bower}, G.~C., {Backer}, D.~C., {Zhao}, J., {Goss}, M., \& {Falcke}, H.
  1999{\natexlab{a}}, \apj, 521, 582

\bibitem[{{Bower} {et~al.}(1999{\natexlab{b}}){Bower}, {Backer}, {Zhao},
  {Goss}, \& {Falcke}}]{Bower1999ApJb}
{Bower}, G.~C., {Backer}, D.~C., {Zhao}, J.-H., {Goss}, M., \& {Falcke}, H.
  1999{\natexlab{b}}, \apj, 521, 582

\bibitem[{{Bower} {et~al.}(1999{\natexlab{c}}){Bower}, {Falcke}, \&
  {Backer}}]{Bower1999ApJc}
{Bower}, G.~C., {Falcke}, H., \& {Backer}, D.~C. 1999{\natexlab{c}}, \apjl,
  523, L29

\bibitem[{{Bower} {et~al.}(2005){Bower}, {Falcke}, {Wright}, \&
  {Backer}}]{Bower2005ApJ}
{Bower}, G.~C., {Falcke}, H., {Wright}, M.~C., \& {Backer}, D.~C. 2005, \apjl,
  618, L29

\bibitem[{{Bower} {et~al.}(2003){Bower}, {Wright}, {Falcke}, \&
  {Backer}}]{Bower2003ApJ}
{Bower}, G.~C., {Wright}, M.~C.~H., {Falcke}, H., \& {Backer}, D.~C. 2003,
  \apj, 588, 331

\bibitem[{{Brunthaler} {et~al.}(2006){Brunthaler}, {Bower}, \&
  {Falcke}}]{Brunthaler2006A&A}
{Brunthaler}, A., {Bower}, G.~C., \& {Falcke}, H. 2006, \aap, 451, 845

\bibitem[{{Brunthaler} {et~al.}(2001){Brunthaler}, {Bower}, {Falcke}, \&
  {Mellon}}]{Brunthaler2001ApJ}
{Brunthaler}, A., {Bower}, G.~C., {Falcke}, H., \& {Mellon}, R.~R. 2001, \apjl,
  560, L123

\bibitem[{{Corbel} \& {Fender}(2002)}]{CorbelFender2002ApJ}
{Corbel}, S. \& {Fender}, R.~P. 2002, \apjl, 573, L35

\bibitem[{{Devereux} {et~al.}(2003){Devereux}, {Ford}, {Tsvetanov}, \&
  {Jacoby}}]{Devereux2003AJ}
{Devereux}, N., {Ford}, H., {Tsvetanov}, Z., \& {Jacoby}, G. 2003, \aj, 125,
  1226

\bibitem[{{Eckart} {et~al.}(2004){Eckart}, {Baganoff}, {Morris}, {Bautz},
  {Brandt}, {Garmire}, {Genzel}, {Ott}, {Ricker}, {Straubmeier}, {Viehmann},
  {Sch{\" o}del}, {Bower}, \& {Goldston}}]{Eckart2004A&A}
{Eckart}, A., {Baganoff}, F.~K., {Morris}, M., {et~al.} 2004, \aap, 427, 1

\bibitem[{{Eckart} {et~al.}(2006){Eckart}, {Baganoff}, {Sch{\"o}del}, {Morris},
  {Genzel}, {Bower}, {Marrone}, {Moran}, {Viehmann}, {Bautz}, {Brandt},
  {Garmire}, {Ott}, {Trippe}, {Ricker}, {Straubmeier}, {Roberts},
  {Yusef-Zadeh}, {Zhao}, \& {Rao}}]{Eckart2006A&A}
{Eckart}, A., {Baganoff}, F.~K., {Sch{\"o}del}, R., {et~al.} 2006, \aap, 450,
  535

\bibitem[{{Falcke}(1996)}]{Falcke1996ApJ}
{Falcke}, H. 1996, \apjl, 464, L67+

\bibitem[{{Falcke} \& {Biermann}(1995)}]{FalckeBiermann1995A&A}
{Falcke}, H. \& {Biermann}, P.~L. 1995, \aap, 293, 665

\bibitem[{{Falcke} {et~al.}(2004){Falcke}, {K{\"o}rding}, \&
  {Markoff}}]{Falcke2004A&A}
{Falcke}, H., {K{\"o}rding}, E., \& {Markoff}, S. 2004, \aap, 414, 895

\bibitem[{{Falcke} \& {Markoff}(2000)}]{FalckeMarkoff2000A&A}
{Falcke}, H. \& {Markoff}, S. 2000, \aap, 362, 113

\bibitem[{{Fender} {et~al.}(2003){Fender}, {Gallo}, \&
  {Jonker}}]{Fender2003MNRAS}
{Fender}, R.~P., {Gallo}, E., \& {Jonker}, P.~G. 2003, \mnras, 343, L99

\bibitem[{{Freedman} {et~al.}(1994){Freedman}, {Hughes}, {Madore}, {Mould},
  {Lee}, {Stetson}, {Kennicutt}, {Turner}, {Ferrarese}, {Ford}, {Graham},
  {Hill}, {Hoessel}, {Huchra}, \& {Illingworth}}]{Freedman1994ApJ}
{Freedman}, W.~L., {Hughes}, S.~M., {Madore}, B.~F., {et~al.} 1994, \apj, 427,
  628

\bibitem[{{Genzel} {et~al.}(2003){Genzel}, {Sch{\" o}del}, {Ott}, {Eckart},
  {Alexander}, {Lacombe}, {Rouan}, \& {Aschenbach}}]{Genzel2003Natur}
{Genzel}, R., {Sch{\" o}del}, R., {Ott}, T., {et~al.} 2003, \nat, 425, 934

\bibitem[{{Ghez} {et~al.}(2004){Ghez}, {Wright}, {Matthews}, {Thompson}, {Le
  Mignant}, {Tanner}, {Hornstein}, {Morris}, {Becklin}, \&
  {Soifer}}]{Ghez2004ApJ}
{Ghez}, A.~M., {Wright}, S.~A., {Matthews}, K., {et~al.} 2004, \apjl, 601, L159

\bibitem[{{Herrnstein} {et~al.}(2004){Herrnstein}, {Zhao}, {Bower}, \&
  {Goss}}]{Herrnstein2004AJ}
{Herrnstein}, R.~M., {Zhao}, J.-H., {Bower}, G.~C., \& {Goss}, W.~M. 2004, \aj,
  127, 3399

\bibitem[{{Ho}(1999)}]{Ho1999ApJ}
{Ho}, L.~C. 1999, \apj, 516, 672

\bibitem[{{Ho} {et~al.}(1996){Ho}, {Filippenko}, \& {Sargent}}]{Ho1996ApJ}
{Ho}, L.~C., {Filippenko}, A.~V., \& {Sargent}, W.~L.~W. 1996, \apj, 462, 183

\bibitem[{{Ho} {et~al.}(1999){Ho}, {van Dyk}, {Pooley}, {Sramek}, \&
  {Weiler}}]{Ho1999AJ}
{Ho}, L.~C., {van Dyk}, S.~D., {Pooley}, G.~G., {Sramek}, R.~A., \& {Weiler},
  K.~W. 1999, \aj, 118, 843

\bibitem[{{Ishisaki} {et~al.}(1996){Ishisaki}, {Makishima}, {Iyomoto},
  {Hayashida}, {Inoue}, {Mitsuda}, {Tanaka}, {Uno}, {Kohmura}, {Mushotzky},
  {Petre}, {Serlemitsos}, \& {Terashima}}]{Ishisaki1996PASJ}
{Ishisaki}, Y., {Makishima}, K., {Iyomoto}, N., {et~al.} 1996, \pasj, 48, 237

\bibitem[{{La Parola} {et~al.}(2004){La Parola}, {Fabbiano}, {Elvis},
  {Nicastro}, {Kim}, \& {Peres}}]{LaParola2004ApJ}
{La Parola}, V., {Fabbiano}, G., {Elvis}, M., {et~al.} 2004, \apj, 601, 831

\bibitem[{{Markoff} {et~al.}(2001{\natexlab{a}}){Markoff}, {Falcke}, \&
  {Fender}}]{Markoff2001A&Ab}
{Markoff}, S., {Falcke}, H., \& {Fender}, R. 2001{\natexlab{a}}, \aap, 372, L25

\bibitem[{{Markoff} {et~al.}(2001{\natexlab{b}}){Markoff}, {Falcke}, {Yuan}, \&
  {Biermann}}]{Markoff2001A&A}
{Markoff}, S., {Falcke}, H., {Yuan}, F., \& {Biermann}, P.~L.
  2001{\natexlab{b}}, \aap, 379, L13

\bibitem[{{Markoff} {et~al.}(2003){Markoff}, {Nowak}, {Corbel}, {Fender}, \&
  {Falcke}}]{Markoff2003A&A}
{Markoff}, S., {Nowak}, M., {Corbel}, S., {Fender}, R., \& {Falcke}, H. 2003,
  \aap, 397, 645

\bibitem[{{Marscher} \& {Gear}(1985)}]{MarscherGear1985ApJ}
{Marscher}, A.~P. \& {Gear}, W.~K. 1985, \apj, 298, 114

\bibitem[{{Melia} \& {Falcke}(2001)}]{Melia2001ARA&A}
{Melia}, F. \& {Falcke}, H. 2001, \araa, 39, 309

\bibitem[{{Miyazaki} {et~al.}(2004){Miyazaki}, {Tsutsumi}, \&
  {Tsuboi}}]{Miyazaki2004ApJ}
{Miyazaki}, A., {Tsutsumi}, T., \& {Tsuboi}, M. 2004, \apjl, 611, L97

\bibitem[{{Narayan} \& {Yi}(1994)}]{NarayanYi1994ApJ}
{Narayan}, R. \& {Yi}, I. 1994, \apjl, 428, L13

\bibitem[{{Page} {et~al.}(2004){Page}, {Soria}, {Zane}, {Wu}, \&
  {Starling}}]{Page2004A&A}
{Page}, M.~J., {Soria}, R., {Zane}, S., {Wu}, K., \& {Starling}, R.~L.~C. 2004,
  \aap, 422, 77

\bibitem[{{Quataert}(2003)}]{Quataert2003ANS}
{Quataert}, E. 2003, Astronomische Nachrichten Supplement, 324, 435

\bibitem[{{Quataert} \& {Gruzinov}(1999)}]{QuataertGruzinov1999ApJ}
{Quataert}, E. \& {Gruzinov}, A. 1999, \apj, 520, 248

\bibitem[{{Reuter} \& {Lesch}(1996)}]{ReuterLesch1996A&A}
{Reuter}, H.-P. \& {Lesch}, H. 1996, \aap, 310, L5

\bibitem[{{Sakamoto} {et~al.}(2001){Sakamoto}, {Fukuda}, {Wada}, \&
  {Habe}}]{Sakamoto2001AJ}
{Sakamoto}, K., {Fukuda}, H., {Wada}, K., \& {Habe}, A. 2001, \aj, 122, 1319

\bibitem[{{Valtaoja} {et~al.}(1992){Valtaoja}, {Terasranta}, {Urpo},
  {Nesterov}, {Lainela}, \& {Valtonen}}]{Valtaoja1992A&A}
{Valtaoja}, E., {Terasranta}, H., {Urpo}, S., {et~al.} 1992, \aap, 254, 71

\bibitem[{{Yuan} {et~al.}(2002){Yuan}, {Markoff}, \& {Falcke}}]{Yuan2002A&A}
{Yuan}, F., {Markoff}, S., \& {Falcke}, H. 2002, \aap, 383, 854

\end{thebibliography}

\end{document}